\documentclass[pre,nofootinbib,twocolumn]{revtex4}
\usepackage{longtable}
\usepackage{amsmath}
\usepackage{graphicx}
\usepackage{amssymb}
\usepackage{epstopdf}
\usepackage{pdfpages}
\usepackage{dcolumn}
\usepackage{hyperref}
\usepackage{bm}

\newcommand{\be}{\begin{equation}}
\newcommand{\ee}{\end{equation}}

\begin{document}

\begin{abstract}
The photoelectron spectrum in the ultra-relativistic limit of tunneling ionization is strongly affected by wave-particle resonance and finite spot-size effects, in contradistinction with the usual assumptions of strong field physics.  Near term laser facilities will access a regime where ionized electrons are abruptly accelerated in the laser propagation direction, such that they stay in phase with the laser fields through a substantial portion of the confocal region.  The final momentum of the electron depends significantly on where in the confocal region it originated.  By manipulating the target and collection geometry, it is possible to obtain low emittance, low energy spread, gigavolt photoelectrons.  Radiation reaction effects play a negligible role in near term scenarios, but become interesting in the multi-exawatt regime.  A significant advance in numerical particle tracking is introduced.
\end{abstract}

\title{Gigavolt bound-free transitions driven by extreme light}

\author{D.F. Gordon, J.P. Palastro, and B. Hafizi}
\affiliation{Beam Physics Branch, Plasma Physics Division}
\maketitle

\section{Introduction}

\begin{figure}
\includegraphics[width=2.5in]{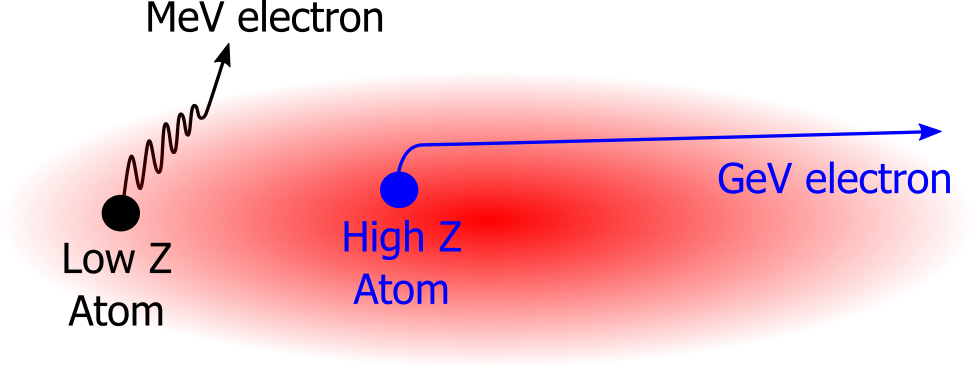}
\caption{Configuration of photoelectron generation.  The red area represents the confocal region of an extreme light laser pulse propagating from left to right.  Electrons ionized from low charge states (black orbit) are spawned outside the extreme field region and are ejected ponderomotively.  Electrons ionized from high charge states (blue orbit) are spawned in an extreme field and can become resonant with the optical wave.  Resonant electrons are forward directed and may obtain GeV energies.}
\label{fig:schematic}
\end{figure}

As laser technology continues to advance, new regimes of laser-matter interaction appear. High-intensity lasers are associated with the threshold for tunneling ionization of outer shell electrons, about $10^{14}$ W$/$cm$^2$.  Ultra-intense lasers are associated with the threshold for relativistic electron dynamics, about $10^{18}$ W$/$cm$^2$.  Extreme light facilities will produce intensities of $10^{22}$ W$/$cm$^2$.  The physics defining this regime is yet to be determined.  This paper points out a significant effect that is likely to be observed: the transition of photoelectron spectra from a ponderomotive to a wave-particle resonance regime.  Fig.~\ref{fig:schematic} illustrates the scenario under consideration.

A fundamental high field process is multi-photon ionization.  The photoelectron spectrum typically ranges from the few eV range, as in above threshold ionization (ATI) \cite{javanainen88,corkum89}, to the few MeV range, as in laser ionization and ponderomotive acceleration (LIPA) \cite{moore99,moore01.lipa}.  The high order limit of multi-photon ionization is tunneling, in which the probability current of the outgoing electron depends only on the instantaneous field \cite{keldysh65,yudin01,popov2004,klaiber13}.  Although this leads to an electron distribution that is initially confined to a small region of phase, during the subsequent motion electrons are exposed to a large number of optical cycles before leaving the confocal region.  A new type of photoelectron spectrum is obtained when the free electron stays confined to a small region of phase at all points in the confocal region, i.e., when it is in phase resonance with the optical wave.  In this paper the laser pulse parameters needed to access this new regime are clearly identified, and it is shown that near-term multi-petawatt laser facilities \cite{ELI-NP} are suitable for an experimental demonstration.

This work makes contact with both the strong field physics of atoms \cite{keldysh65,moore99,moore01.lipa,yudin01,popov2004,klaiber13}, and the acceleration of electrons by electromagnetic fields in free space \cite{palmer87,kimura95,hartemann98,wang01,gupta07,cline13}. Of the many articles pertaining to strong field physics, those concerning the photoelectron spectra produced in the LIPA scheme \cite{moore99,moore01.lipa} are most closely related to this work.  The process described herein can be thought of as the extreme-light limit of LIPA, hereafter abbreviated xLIPA.  In LIPA, electrons originate from a tenuous gas of moderately heavy elements.  Upon focusing an ultra-intense laser pulse into the gas, electrons are tunnel ionized and accelerated.  When the ionization potential is matched to the laser power and focusing, such that the electrons are ionized only when they are exposed to the peak laser intensity, free electrons are spawned in a large ponderomotive potential, and gain high energy.  In the case of xLIPA, the ionization potential has to be matched in a similar way, but the acceleration is no longer ponderomotive.  Instead, electrons stripped from atoms with certain favorable initial positions are brought abruptly to the speed of light and gain further energy as they stay in nearly the same phase of the laser field for extended periods of time. 

Of the many articles pertaining to free space acceleration, those concerning the ``Capture and Accelerate Scenario'' (CAS) \cite{wang01,cline13} are most closely related to this work.  In this scheme, electrons are externally injected into a laser focus, and are abruptly accelerated to the speed of light.  CAS differs from xLIPA primarily in the nature of the initial conditions.  Namely, in xLIPA the intial conditions are tied to the strong field physics of atoms, whereas in CAS they are engineered using a conventional accelerator.  In practical terms, xLIPA accesses the CAS regime without the need for a conventional accelerator.

\section{Overview of Results}

\begin{figure}
\includegraphics[width=3.5in]{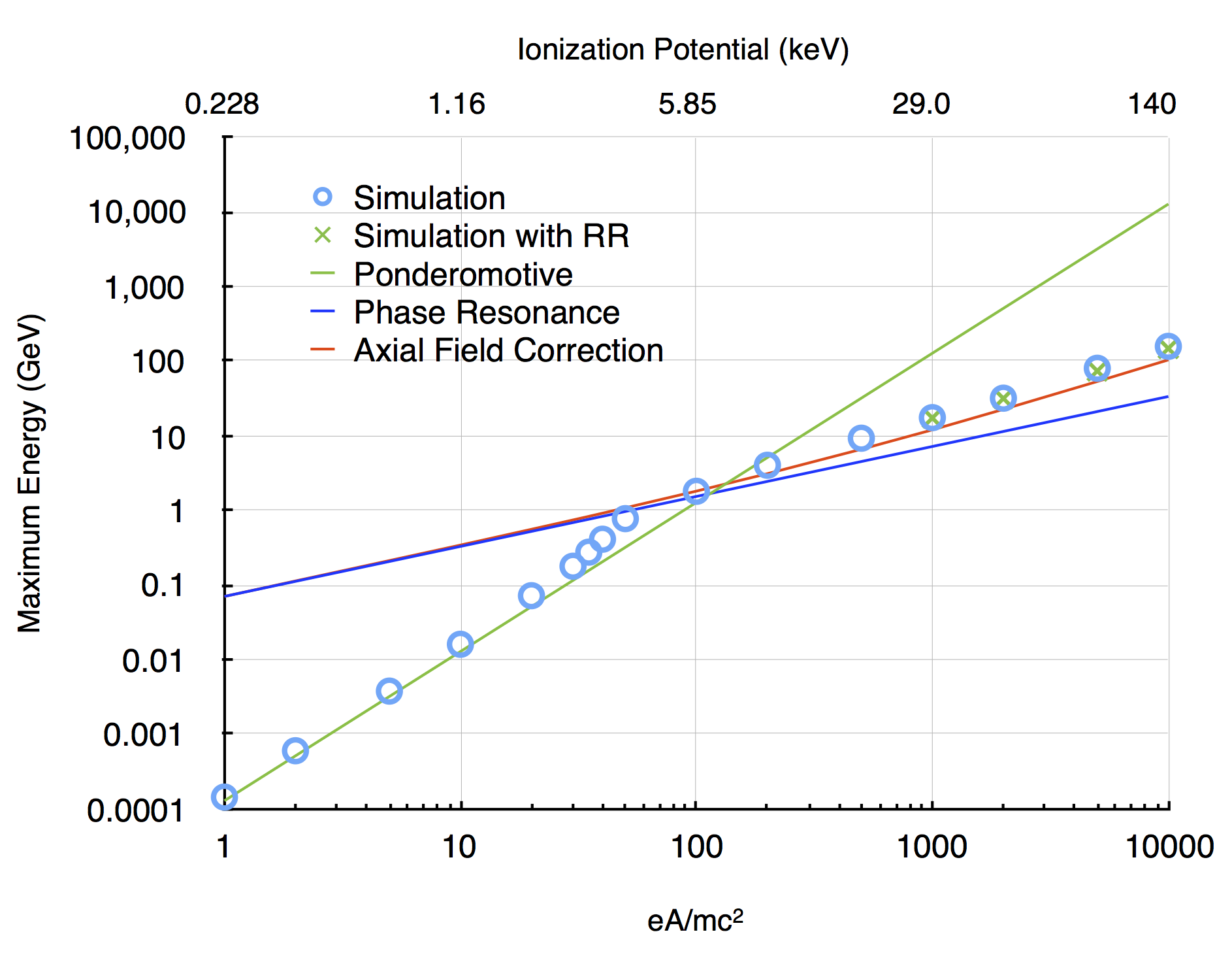}
\caption{Maximum energy in the photoelectron spectrum for a 0.8 $\mu$m wavelength, 20 fs pulse, focused to a 5 $\mu$m spot (pulse metrics are referenced to $1/e$ of the field).}
\label{fig:ascan}
\end{figure}

The primary result of this paper in shown in Fig.~\ref{fig:ascan}, which plots the maximum photoelectron energy as a function of the normalized vector potential, $a = eA/mc^2$.  The parameter $a$ characterizes the importance of relativistic effects, i.e., when $a \lesssim 1$ the free electron dynamics are weakly-relativistic, while when $a \gg 1$ they are ultra-relativistic.  Each value of $a$ is paired with a particular ionization potential, $U_{\rm ion}$, shown on the upper horizontal axis.  The $U_{\rm ion}$ values are chosen so that the corresponding $a$ is twice the threshold value for tunneling ionization.  The plot shows simulated data with and without radiation reaction (RR), along with three theoretical curves for comparison.  The numerical methods used for the simulation are in the appendix, and the derivation of the theoretical curves is discussed below.

Fig.~\ref{fig:ascan} illustrates several important points.  First, two regimes of photoelectron generation are clearly suggested, one for $a \lesssim 100$ and one for $a \gtrsim 100$.  The simulation results for $a\leq 10$ are well matched by the theoretical curve marked ``ponderomotive,'' which is just the prediction of LIPA.  The phase resonance model, which is derived below, gives two different curves depending on whether axial field components are kept or dropped in the analysis.  While both of these curves fit the data for $a\gtrsim 100$ better than the LIPA model, the one that accounts for the axial field components is more accurate.  In simulations where the axial field is artificially suppressed, the maximum energy gain is indeed reduced as the theoretical curves indicate.

Perhaps the most interesting features of Fig.~\ref{fig:ascan} are the persistence of ponderomotive scaling for $a \gg 1$, and the even steeper slope for $30<a<50$.  These features illustrate that the transition between ponderomotive and phase resonance acceleration is characterized by a rapid increase in energy, rather than the onset of a limiting factor in the scaling law.  It is this feature that makes xLIPA interesting not only from the strong field physics point of view, but also from the point of view of free space acceleration of electrons.

Finally, RR has only a small effect on the maximum photoelectron energy, even for highly speculative values of $a$.  However, the effect on the momentum distribution is interesting, as is elaborated upon below.  RR effects are limited to the multi-exawatt regime, and should play no role in near-term experiments.

\section{Analysis of Phase Resonance}
\label{sec:analysis}

In this section, useful estimates are derived for the maximum energy obtained by a free electron that is abruptly introduced into an extreme laser field.  The analysis is based on the expectation that when an inner shell electron is tunnel-ionized, it is accelerated abruptly to the speed of light.  In this limit, the motion is nearly parallel to the wavevector of the radiation, and the phase of the particle in the radiation field can be regarded as constant.  The primary constraint is that the interaction is limited to regions where the irradiance is high and the phase velocity is close to $c$.   This corresponds to the two regions just outside the confocal region.  That is, far from the confocal region the irradiance is too low, but inside the confocal region the phase velocity is too high.

There is an exact solution for the motion of a charged particle in any radiation field of the form $F_{\mu\nu}(k_\mu x^\mu)$, where $k_\mu$ is the four dimensional wavevector of the radiation, and $x_\mu$ is the spacetime coordinate. Without loss of generality, let $k_1 = k_2 = 0$.  Then
\be
\Upsilon \equiv u_0 - u_3
\ee
is invariant \cite{popov2004,klaiber13}, where $u_\mu$ is the four dimensional velocity of the particle\footnote{In particular, $u = (\gamma,\gamma\beta_1,\gamma\beta_2,\gamma\beta_3)$, where $c\beta_i$ are the components of the three dimensional velocity, and $\gamma = (1 - \beta^2)^{-1/2}$}.  Combining this with the identity $u_\mu u^\mu = 1$, one can show that in the high energy limit $u_3 \gg \{u_1,u_2\}$.  In other words, the particle moves predominatly in the ``forward'' direction, i.e., parallel to the radiation wavevector.

Consider the lowest order, linearly polarized, Hermite-Gaussian laser mode, with $x_1$ the polarization direction and $x_3$ the propagation direction.  The equations of motion for a perfectly resonant particle can be integrated most conveniently in the case $x_2 = 0$, so that the axial magnetic field vanishes.  As will be shown below, this is the most interesting case for high energy photoelectron production.  In matrix form, the equations of motion are
\be
\frac{dx}{ds} = cu
\ee
\be
\frac{du}{ds} = \Omega u
\ee
where $x(s)$ is the world line of the particle, $u(s)$ is the four-velocity, and $\Omega = a \omega F$.  The parameter $s$ is the proper time, $a = qE/mc\omega$, $E$ is the electric field, $q$ is the charge of the particle, $m$ is the mass, and $\omega$ is the frequency of the radiation.  Using the coordinate system described above, the matrix $F$ is
\setlength\arraycolsep{10pt}
\be
F = 
\left(
\begin{array}{cccc}
0  &1   & 0  & \epsilon \\
1  & 0  & 0 & -1 \\
0 &  0 & 0 & 0 \\
\epsilon & 1 & 0 & 0  
\end{array}
\right)
\ee
Here, $\epsilon$ is the ratio of axial to transverse electric field, which need not be small.  Define a phase resonant particle as one for which $\Omega$ is slowly varying on $x(s)$.  Such particles have $x(s)$ confined to the intersection of two regions, one being a neighborhood about a hypersurface of constant $\epsilon$, and the other a neighborhood about a hypersurface of constant phase.  In the first approximation, $\Omega$ is constant, and the solution of the velocity equation is
\be
u(s) = e^{\Omega s} u(0)
\ee
The matrix exponential $e^{\Omega s}$ is tractable, and can be easily computed using symbolic math software.  Let
\be
\Lambda(s) \equiv e^{\Omega s}
\ee
It can be shown that $\Lambda$ is a Lorentz transformation, i.e., $\Lambda^T g \Lambda = g$, where $T$ indicates the transpose, and $g = {\rm diag}(1,-1,-1,-1)$.  Of particular interest is the initial condition $u(0) = (1,0,0,0)^T$, which according to most theories holds for an electron at the moment of ionization, at least when the atomic number satisfies $Z \ll 137$.  In this case,
\be
u(s) = \frac{1}{\epsilon^2}\left(
\begin{array}{cccc}
\cosh\epsilon\sigma-1+\epsilon^2\cosh\epsilon\sigma \\
\epsilon-\epsilon\cosh\epsilon\sigma + \epsilon\sinh\epsilon\sigma \\
0  \\
\cosh\epsilon\sigma - 1 + \epsilon^2\sinh\epsilon\sigma 
\end{array}
\right)
\ee
where $\sigma(s) = a\omega s$.  This gives $\Upsilon = \cosh\epsilon\sigma - \sinh\epsilon\sigma$, so that the invariance of $\Upsilon$ is broken.  In the plane wave limit ($\epsilon \rightarrow 0$) the particle momentum is
\be
u(s) = \left(
\begin{array}{cccc}
1 + \sigma^2/2 \\
\sigma \\
0  \\
\sigma^2/2 
\end{array}
\right)
\ee
and the invariance of $\Upsilon$ is restored, i.e., $\Upsilon = 1$.  Moreover, when $\sigma \gg 1$, the momentum is predominantly in the forward direction, i.e., $u_3 \gg u_1$.  Assuming $a \gg 1$, this requires that $\omega s$ be at least of order unity, i.e., the time elapsed according to a clock moving with the particle should read at least one laser period, as measured by a lab frame clock.  This does not necessarily violate the assumption that the particle should stay in phase, since the two clocks may keep very different time.

In order to estimate the maximum energy gain, values for $s$ and $\epsilon$ are needed.  The distance traversed by the particle is
\be
\begin{aligned}
x_3(s) = &c\int u_3(s) ds \\
= &\frac{\lambda}{2\pi\epsilon^3 a}\left( \sinh\epsilon\sigma - \epsilon\sigma + \epsilon^2\cosh\epsilon\sigma \right)
\end{aligned}
\ee
where $\lambda = 2\pi c/\omega$.  The maximum energy gain is $u_0(s')$, where $s'$ is the solution of $x_3(s') = z_R$.  Here, $z_R = \pi r_0^2/\lambda$ is the Rayleigh length, with $r_0$ the radius of the beam waist.  A characteristic value for $\epsilon$ is obtained by evaluating the field at the point $(r_0,0,z_R)$, which gives $\epsilon^2 = \lambda/4\pi z_R$.

A simple formula for the maximum energy gain can be obtained by using the plane wave estimate for the momentum.  In this case, the proper time is related to $x_3$ by
\be
s = \frac{(6x_3)^{1/3}}{c(ak)^{2/3}}
\ee
so that the energy can be put explicitly in terms of $x_3$:
\be
u_0(x_3) = 1 + \frac{1}{2}\left(6ak x_3\right)^{2/3}
\ee
Substituting the Rayleigh length for $x_3$ gives.
\be
\label{eq:u0}
u_{0,{\rm max}} = 1 + \frac{1}{2}\left(\frac{\pi r_0}{\lambda}\right)^{4/3} (12a)^{2/3} 
\ee
As an example, a 10 PW laser pulse, with $\lambda = 0.8$ $\mu$m, focused to $r_0 = 5$ $\mu$m, gives $a \approx 100$, and $u_{0,{\rm max}}mc^2 \approx 1.5$ GeV.  Comparison of the prediction accounting for the axial field with that of Eq. (\ref{eq:u0}) is in Fig.~\ref{fig:ascan}.  The axial field correction curve is obtained by solving for $s'$ numerically.

In the first approximation, there is no advantage in varying the laser wavelength.  To see this, rewrite Eq.~(\ref{eq:u0}) as
\be
u_{0,{\rm max}} = 1 +2 \left(\frac{3\pi r_0}{\lambda}\right)^{2/3} \left(\frac{Pr_e}{mc^3}\right)^{1/3} 
\ee
where $P$ is the laser power and $r_e$ is the classical electron radius.  The wavelength appears only in the combination $r_0/\lambda$, which is fixed by the focusing geometry.

The essential element in the forgoing analysis is the assumption that upon ionization into an extreme field, an electron can be accelerated to nearly the speed of light in a fraction of an optical cycle.  This requires that the ionization potential be large enough so that the electron is held in position by its parent ion until it is exposed to ultra-relativisitic intensity, but not so large that ionization becomes highly improbable.

\section{Two-step Ionization Model}
\label{sec:twostep}

An elegant, but incomplete, calculation, of the momentum distribution of relativistic tunnel-ionized electrons, can be carried out in the strong field approximation (SFA) \cite{reiss92}.  Recently, the SFA was extended to account for the Coulomb field of the residual ion by means of a quasi-classical correction factor \cite{klaiber13}.  An alternative approach is the imaginary time method (ITM) \cite{popov2004}.  The SFA is perhaps more suitable for the purposes of the present paper, due to the fact that it utilizes the S-matrix, which gives directly the primary object of interest, the momentum distribution of the ionized electron.  For this reason, the following discussion emphasizes the SFA over the ITM.

A major claim of this paper is that the assumptions typically utilized in practical SFA analyses fail in the extreme light regime.  Foremost among these, are the restrictions on the form of the laser field, which is invariably taken to have perfectly planar phase fronts, and slowly varying envelope.  In order to remove these assumptions, the two-step model suggested by Corkum \cite{corkum89} is employed.  In the first step, a free electron is spawned at a random time, chosen so that the probability of ionization during any time interval is consistent with some rate law (see appendix).  In the second step, the free electron evolves according to the classical laws of motion until it leaves the interaction region.  Carrying out a large number of trials gives what may be termed a classical S-matrix, $S_{\bf rp}$, where $|S_{\bf rp}|^2$ is the probability that an ion with spatial coordinate ${\bf r}$ will produce a photoelectron with momentum ${\bf p}$.  This dependence on the ion coordinates, absent in the SFA, comes about because of finite spot-size effects.

\begin{figure}
\includegraphics[width=3.0in]{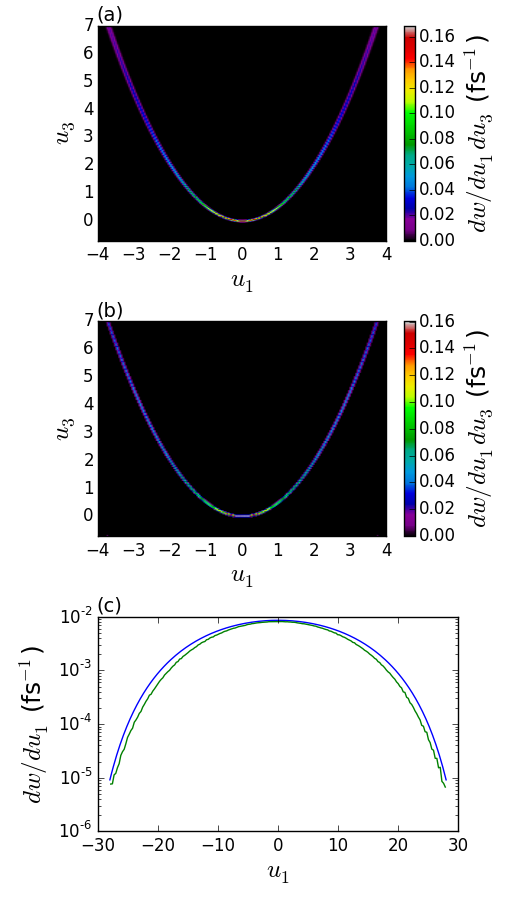}
\caption{Benchmarking the two-step model in the plane wave limit, with $a=36$, $Z=18$, and $\tau = 20$ fs, by comparing the differential ionization rate with the SFA prediction.  Panel (a) is the analytical SFA formula, (b) is the numerical two-step model, and (c) projects out $u_3$ to allow for better quantitative comparison.  In (c), the blue curve is the SFA, while the green curve is the two-step model.}
\label{fig:twostep}
\end{figure}

The two-step model requires a rate law, which in this work is the Coulomb corrected, dressed, SFA of Klaiber et al. \cite{klaiber13}.  This creates the difficulty that a subset of the information contained within the SFA (the rate law) is used to obtain a result that differs from the total information contained within the SFA (the momentum distribution).  In order for this to make sense, it is required that the results from the two-step model should not depend strongly on which rate law is used.  To this end, we verified that the momentum distributions derived from the two-step model are nearly indistinguishable whether the ITM, the dressed SFA, or the undressed SFA, are employed in the rate law.

Confirmation that the two-step model agrees with the SFA, in the limit of a plane wave, is given in Fig.~\ref{fig:twostep}.  The peak normalized vector potential of the laser is $a=36$, and the pulse width is $20$ fs.  The ion under consideration is hydrogen-like argon, $Z=18$.  Various projections of the differential ionization rate are plotted using both the SFA and the two-step model.  In computing a differential ionization rate from the two-step model output, the $S$-matrix probabilities have to be divided by a suitable time interval.  The time interval used to generate Fig.~\ref{fig:twostep} is the width of the ionization current pulse envelope.  This is determined by monitoring the number of ionization events per unit time, integrated over all trials.

The parabolic shape of the distribution in the $u_1$-$u_3$ plane, which follows from the invariance of $\Upsilon$, is indistinguishable from one case to the other.  The relative density distribution along the parabola is also very similar.  Comparison of the absolute scale in the $u_1$-$u_3$ plane is not meaningful, because strictly, the two-step model distribution contains a factor $\delta(\Upsilon-1)$, whereas the SFA allows for small deviations about $\Upsilon = 1$.  The value of $dw/du_1du_3$ is rendered finite in Fig.~\ref{fig:twostep}(b) by applying a smoothing filter.  In order to make a comparison where the absolute scale has meaning, Fig.~\ref{fig:twostep}(c) shows the differential ionization rate projected onto the $u_1$ axis.

Fig.~\ref{fig:twostep} establishes that the two-step model is capable of generating a momentum distribution that closely approximates the SFA in the plane wave limit.  An important characteristic of the plane wave case is that the initial position of the parent ion does not affect the final momentum in any way.  The two-step model allows finite spot-size illumination to be considered, and the effect of the ion coordinates to be determined.

\section{Momentum Distributions}

Consider a system comprised of many ions distributed in space.  Assuming that correlations among photoelectrons are negligible, the total momentum distribution is
\be
\label{eq:fp}
f({\bf p}) = \sum_i \left| S_{{\bf r}_i{\bf p}} \right|^2
\ee
where $i$ indexes each ion. By loading a large number of ions into any given simulation, an ensemble of ion coordinates and ionization times is created, and $f({\bf p})$ is generated by binning the whole set of photoelectrons in momentum space.

\begin{figure}
\includegraphics[width=3.0in]{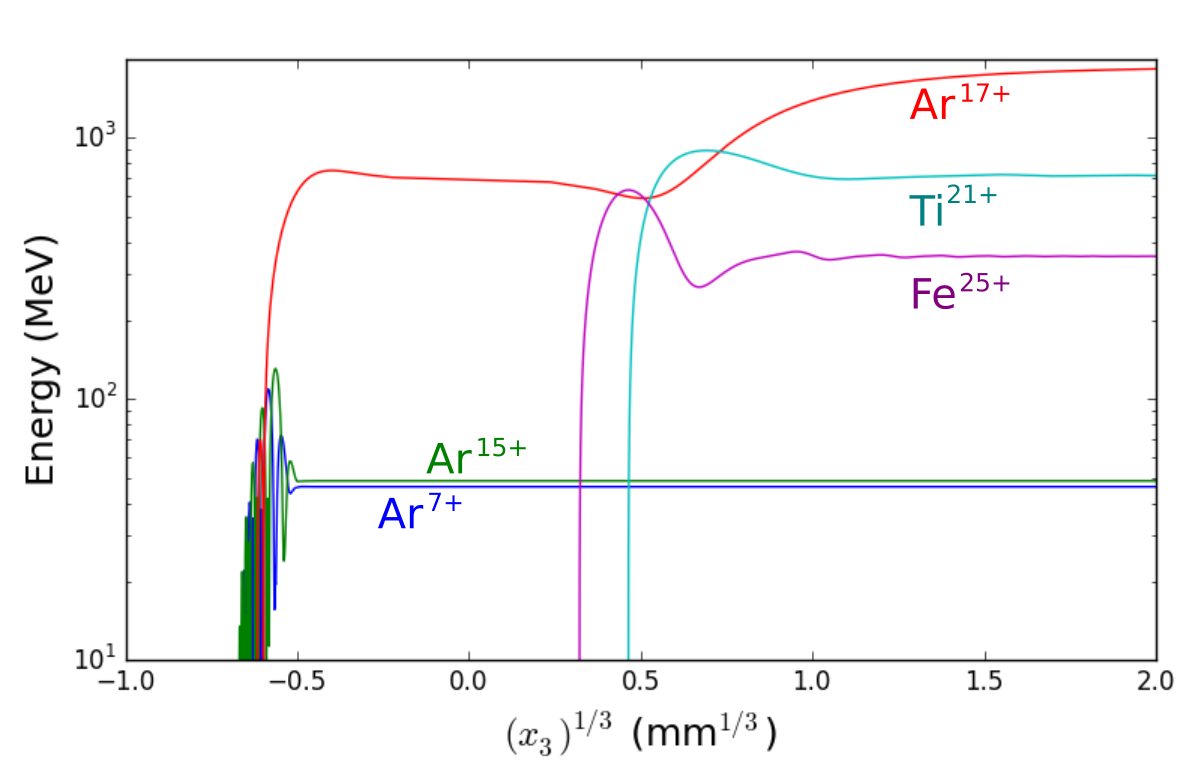}
\caption{Representative orbits of photoelectrons for various parent ions.  Electrons from the $M$ or $L$ shell exhibit ponderomotive acceleration, while electrons from the $K$ shell exhibit phase resonance.  The fractional power scale on the horizontal axis gives the effect of a log scale, while allowing for a transition through the origin.}
\label{fig:orbits}
\end{figure}

Consider laser parameters $a=100$, $\lambda = 0.8$ $\mu$m, $r_0 = 5$ $\mu$m, and $\tau = 20$ fs, which are chosen to be achievable by near-term 10 PW laser systems.  Representative photoelectron orbits from the two-step model are shown in Fig.~\ref{fig:orbits}, for various parent ions.  The optimum energy gain is obtained for Ar$^{17+}$.  The lower charge states of argon are ionized too early in the foot of the pulse, and are ponderomotively ejected before phase resonance can be reached.  The case of Ti$^{21+}$ is similar to that of Ar$^{17+}$, while that of Fe$^{25+}$ gives more limited energy gain.  The limitation in the case of Fe$^{25+}$ comes about because the photoelectrons tend to be spawned too near the origin, which is not the ideal starting location.

\begin{figure}
\includegraphics[width=2.5in]{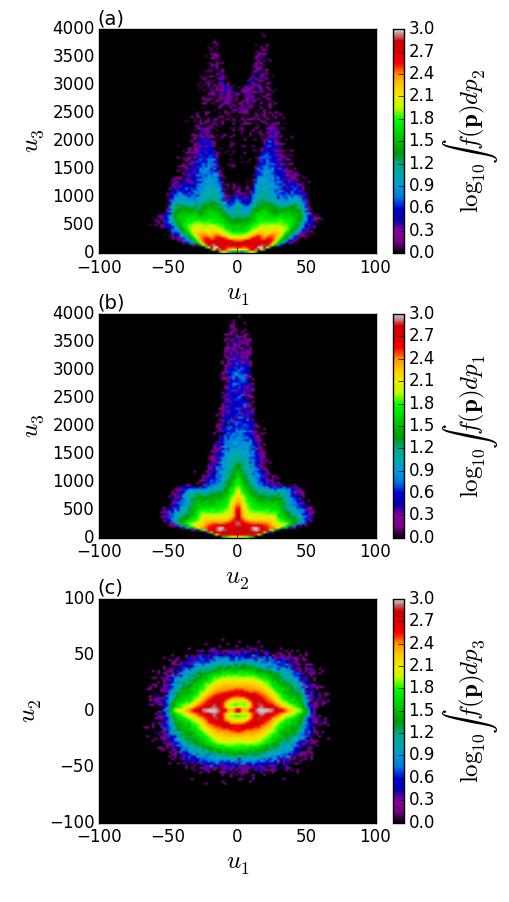}
\caption{Projections of the photoelectron momentum distribution, $f({\bf p})$, for an ensemble of initial conditions corresponding to uniformly distributed Ar$^{17+}$ ions.}
\label{fig:fp}
\end{figure}

\begin{figure}
\includegraphics[width=2.75in]{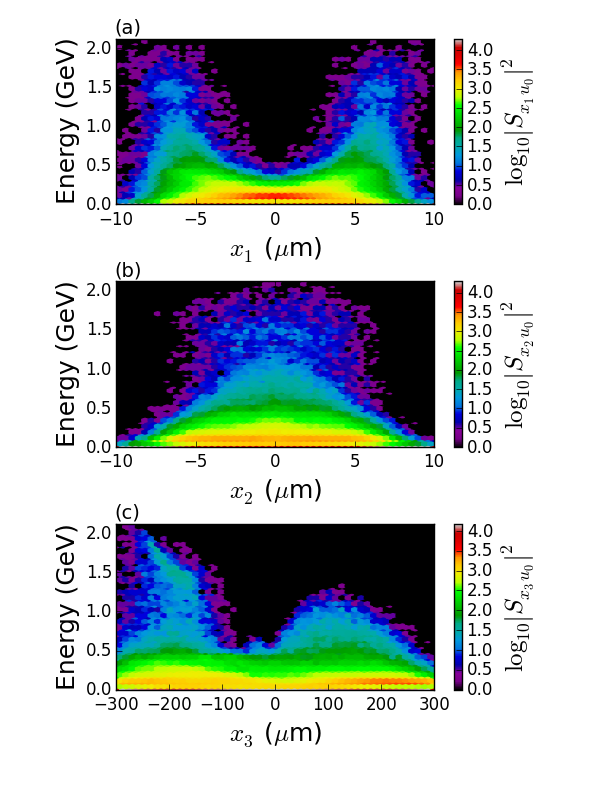}
\caption{Classical S-matrix projections for Ar$^{17+}$.  The correlation of final energy with initial polarization coordinate is in (a), with initial cross-polarization coordinate is in (b), and with initial axial coordinate is in (c).}
\label{fig:SPW}
\end{figure}

The momentum distribution, $f({\bf p})$, of photoelectrons extracted from a uniform distribution of Ar$^{17+}$ ions, is shown in Fig.~\ref{fig:fp}.  Comparison of Fig.~\ref{fig:fp}(a) with Figs.~\ref{fig:twostep}(a) or (b) reveals that finite spot size effects lead to a profound change in the photoelectron spectrum.  This can be easily understood in terms of the fact that in the plane wave case, there is no possibility of any electron staying confined to a small region of phase throughout the interaction.  The interaction stops only after all phases of the plane wave pass by.  When the spot size is finite, the interaction is terminated when the electron exits the confocal region.  In the latter case, phase resonance is possible, and higher energies are obtained.

Projections of the classical S-matrix for Ar$^{17+}$, correlating final energy with initial coordinate, are shown in Fig.~\ref{fig:SPW}.  In any given projection, there is a large low energy population, and only a small number of high energy particles.  The highest energy particles originate roughly from the coordinates $x_1 \approx \pm r_0$, $x_2 \approx 0$, and $x_3 \approx - 2z_R$.  This suggests that a high quality, high energy beam, might be obtained by localizing the parent ions to a small neighborhood about one or both of these two points.  If a gas target is used, some advantage might be gained by focusing into a plasma channel or lens \cite{hubbard02,katzir09}, so that the highest ion density is off-axis.  Alternatively, one might consider suspending titanium nanoparticles at the appropriate points in the laser focus.  Highly accurate positioning and alignment would be needed in either case, and pre-pulses would have to be strongly extinguished.

\begin{figure}
\includegraphics[width=3.25in]{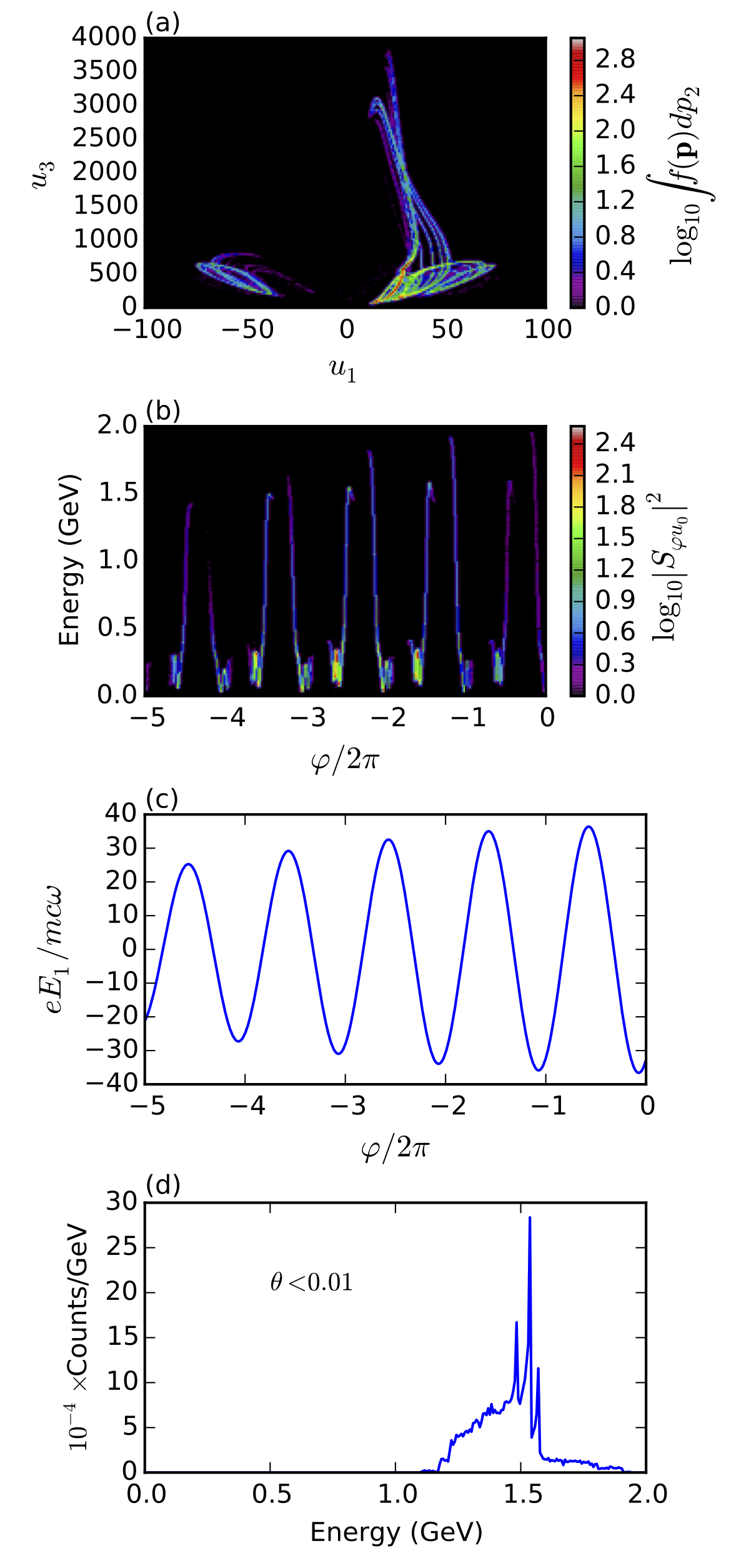}
\caption{Momentum distribution from a single, well positioned, Ar$^{17+}$ ion.  The $u_1$-$u_3$ momentum projection is in (a), and the $\varphi$-$u_0$ $S$-matrix projection is in (b).  The value of $E_1$, evaluated at the position of the ion, is shown in (c) as a function of $\varphi$.  The energy spectrum for photoelectrons inside a 10 milliradian cone angle is in (d).}
\label{fig:fpsingle}
\end{figure}

Leaving these practical matters in abeyance, consider the photoelectron distribution from a single Ar$^{17+}$ ion positioned near the optimal coordinate.  This is computed using $10^6$ trials of the two-step model, with ion coordinate ${\bf x} = (r_0,0,-2z_R)$.  The results are illustrated in Fig.~\ref{fig:fpsingle}.  The $u_1$-$u_3$ projection, shown in Fig.~\ref{fig:fpsingle}(a), shows a clear preference for positive $u_1$ values, evidently due to the choice of positive $x_1$.  The fine structure exhibits discrete filamentary structures, including both open and closed loops.  This structure can be understood by inspecting Fig.~\ref{fig:fpsingle}(b), which shows the $S$-matrix projected into the plane of ionization phase and energy.  The ionization phase is defined as $\varphi = \omega(t_c - x_3/c)$, where $t_c$ is the time of ionization (see appendix).  The transverse electric field component, evaluated at the point ${\bf x}$, is plotted as a function of $\varphi$ in Fig.~\ref{fig:fpsingle}(c).  Because of finite spot-size and pulse-width effects, the peaks occur at non-integral values of $\varphi/\pi$.  Cross-referencing Figs.~\ref{fig:fpsingle}(a), (b), and (c) makes it apparent that the closed loops in (a) are generated by particles ionized near the peaks of the electric field.  The open loops are generated either just before the electric field reaches its peak negative value, or just after it reaches its peak positive value.  The corresponding electrons are situated to stay in phase resonance for the longest time, and therefore reach the highest energies.    Finally, the energy spectrum within a cone angle $\theta = 0.01$ radians is shown in Fig.~\ref{fig:fpsingle}(d).  The three narrow peaks around $1.5$ GeV are associated with the resonant electrons that originate in positive electric fields.  The broad, higher energy tail is due to the resonant electrons that originate in negative electric fields.  

\section{Radiation Reaction Effects}

A fundamental problem in the strong field physics of atomic systems, which seems to have been left unexplored up until now, is the effect of radiation reaction (RR) on the spectrum of tunnel ionized electrons.  RR is the force acting on an electron due to its own fields.  This force is normally negligible, only becoming significant for very large fields or electron energies.  With laser fields reaching ever more stupendous values, the problem of RR is attracting renewed interest \cite{rohrlich02,hartemann05,dipiazza09,thomas12,kumar13,kravets13}.  The fields needed to bring the photoelectron dynamics into the RR dominated regime are far in excess of what is likely to be achievable in the near future.  This can be understood in terms of the fact that high energy photoelectrons always co-propagate with the optical wave, whereas RR is encouraged by counter-propagating geometries.

\begin{figure}
\includegraphics[width=3.5in]{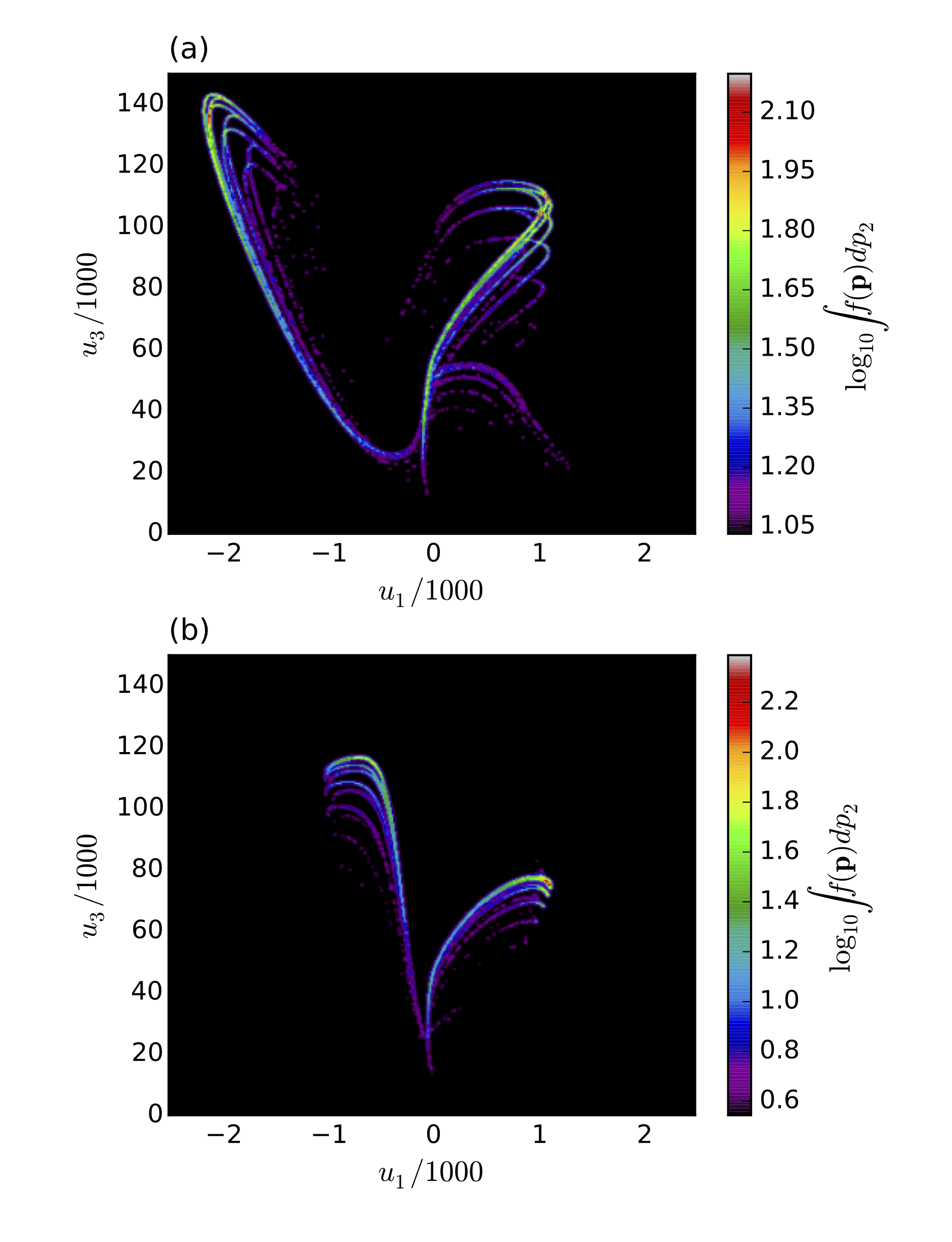}
\caption{Momentum distribution from a single, well positioned, Au$^{78+}$ ion, exposed to an irradiance of $2\times 10^{26}$ W$/$cm$^2$.  The case neglecting RR is in (a), and the case with the exact Landau and Lifshitz formula for RR is in (b).}
\label{fig:RR}
\end{figure}

The numerical methods used here incorporate the exact Landau and Lifshitz formula for RR (see appendix).  Significant effects of RR on the photoelectron spectrum, given the focal geometry considered throughout this paper, begin to appear for $a=5000$, corresponding to about 25 exawatts of laser power.  Such a laser field, according to the SFA, is sufficient to fully strip gold ($Z=79$).  Consider placement of an Au$^{78+}$ ion at the same location in the confocal region considered above, ${\bf x} = (r_0,0,-2z_R)$.  The $u_1$-$u_3$ momentum distributions with and without RR are displayed in Fig.~\ref{fig:RR}.  In this extreme case, the charge is peaked at the highest energies, even without any angular selection.  Either with or without RR, there are two main peaks in the energy spectrum.  The peaks in the spectrum with RR are lower energy, but narrower.  The angular spread in the case with RR is also reduced.  The overall shape in the two cases is obviously quite different.  Developing a full understanding of these effects promises to be a rich area of research.

\section{Conclusions}

Multi-petawatt lasers access a new regime of strong field physics and free space acceleration, which appears when the normalized vector potential $a \gtrsim 100$.  The electric field in the confocal region is sufficient to fully strip moderately heavy atoms such as argon or titanium via tunneling ionization.  The K-shell electrons are released into such a large electric field, they are accelerated to the speed of light in a fraction of an optical period, and are deflected into the laser propagation direction by the magnetic field.  Some of these electrons are accelerated to gigavolt energies due to phase resonance, i.e., they stay near the same optical phase throughout a substantial portion of the confocal region. The resulting photoelectron momentum distributions have unique features, which depart significantly from the usual parabolic form predicted by the SFA and other analyses.  For atoms loaded into a small spatial region, the momentum distribution has discrete features that can be related to sub-cycle bursts of ionization current.

\section{Acknowledgements}

This work was supported by the NRL 6.1 Base Program.  We have enjoyed fruitful discussions touching this work with many colleagues, including A. Ting, D. Kaganovich, M.H. Helle, R.F. Hubbard, L. Johnson, J.R. Pe${\rm \tilde{n}}$ano, J. Giuliani, J. Davis, A. Zigler, A.S. Landsman, A. Noble, K. Krushelnick, A. Maksimchuk, E. Chowdhury,  D. Schumacher, and E. Turcu.

\section*{Appendix A: Extreme Field Particle Pusher}

Ordinary particle pushers are ineffective in extreme fields.  In Ref.~\cite{gordon15.NRLMR}, a covariant pusher with an adaptive time step is used to maintain accuracy.  Here, the covariant form is retained, but an important improvement is made which eliminates the need for explicit time step adjustment, and increases computational efficiency by orders of magnitude.  The treatment of radiation reaction is identical to the one in \cite{gordon15.NRLMR}.

The exact operator of momentum evolution is a Lorentz transformation.  Typically one approximates this as some sequence of boosts and rotations \cite{boris70,verboncoeur05,gordon15.NRLMR}.  Due to the fact that these are not commuting operators, an error is introduced, which is an increasing function of the field strength.  To avoid this problem, a compact form of the complete Lorentz transformation is needed.

The pusher used here advances particles in proper time, and takes a manifestly covariant form.  One useful feature of a proper time advance is that a constant proper time step leads to a constant phase step for any particle in a plane wave, no matter its energy or direction.  In fact, requiring that the phase step satisfy $\Delta \varphi \ll 2\pi$ leads to
\be
\label{eq:cond1}
\Delta s \ll \frac{2\pi}{\omega\Upsilon}
\ee
where $\Upsilon$ is a constant of the motion.  For a particle initially at rest, $\Upsilon = 1$, and the proper time step appropriate for any particle is the same as the lab frame time step appropriate for a non-relativistic particle.

The condition (\ref{eq:cond1}) is sufficient provided the momentum advance is exact in a uniform, constant field.  This requirement is satisfied by the expression $u(s+\Delta s/2) = \Lambda(s,\Delta s)u(s-\Delta s/2)$, where $s$ is the proper time, $\Delta s$ is the time step, and
\be
\Lambda(s,\Delta s) = \exp\left(qF(s)\Delta s/mc\right)
\ee
Here, $F(s)$ is the antisymmetric field tensor evaluated on the world line $x(s)$.  To avoid the appearance of onerously complicated expressions, it is convenient to first rotate the coordinate system so that there is only one component of ${\bf B}$, apply the matrix exponential, and finally restore the coordinate system.  Specifically,
\be
\Lambda(s,\Delta s) = {\sf T}^{-1}({\bf B})\Lambda_0(s,\Delta s){\sf T}({\bf B})
\ee
where ${\sf T}({\bf B})$ aligns the magnetic field with one of the basis vectors.  For definiteness let this be the basis vector in the 1-direction, ${\bf e}_1$.  Then, ${\sf T}({\bf B})$ is any rotation matrix satisfying ${\sf T}({\bf B})\cdot{\bf B} = |{\bf B}|{\bf e}_1$.  The matrix exponential is conveniently split into symmetric and antisymmetric parts, such that $\Lambda_0 = \Lambda_S + \Lambda_A$, where
\begin{widetext}
\be
\Lambda_S = \Xi_+ + \frac{1}{2\alpha}\left(
\begin{array}{cccc}
2F\Xi_-  & 2\sqrt{2}\epsilon_1(F\Sigma_-+\alpha\Sigma_+) & \sqrt{2}\epsilon_2(\alpha_+^2\sigma_+ + \alpha_-^2\sigma_-)  & \sqrt{2}\epsilon_3(\alpha_+^2\sigma_+ + \alpha_-^2\sigma_-) \\
\cdot  & -2G\Xi_-  & 4\epsilon_1\epsilon_2\Xi_- & 4\epsilon_1\epsilon_3\Xi_- \\
\cdot &  \cdot  & -2H\Xi_- & 4\epsilon_2\epsilon_3\Xi_- \\
\cdot & \cdot & \cdot  & -2J\Xi_- 
\end{array}
\right)
\ee
\be
\Lambda_A = \frac{\sqrt{2}b}{\alpha}\left(
\begin{array}{cccc}
0 & 0 & -\sqrt{2}\epsilon_3\Xi_- & \sqrt{2}\epsilon_2\Xi_- \\
0  & 0  & -2\epsilon_1\epsilon_3\Sigma_- & 2\epsilon_1\epsilon_2\Sigma_- \\
\cdot &  \cdot & 0 & G\Sigma_- + \alpha\Sigma_+ \\
\cdot & \cdot & \cdot & 0  
\end{array}
\right)
\ee
\end{widetext}
Here, dotted entries are used to emphasize that the corresponding matrix elements can be obtained from symmetry.  The several variables appearing in the matrix exponential are defined in Table~\ref{tab:matrixexp}.  In programming the matrix, care must be taken to prevent floating point exceptions in regions where $\alpha$ or $\alpha_\pm$ vanish, such as in field free regions, or ideal plane waves.  Imposing a miniscule uniform field is one simple solution.  Also, some advantage might be gained by updating one of the components of $u$ via the identity $u^Tgu = 1$, rather than by using the corresponding row of $\Lambda_0$.  In this work $u_0$ is so updated.

\begin{table}
\caption{Variables Appearing in $\Lambda_0$}
\label{tab:matrixexp}
\begin{center}
\begin{tabular}{lr}
\hline
Variable & Definition \\
\hline\hline
$\epsilon_i$ & $({\sf T}{\bf E})\cdot{\bf e}_i$ \\
$b$ & $({\sf T}{\bf B})\cdot {\bf e}_1$ \\
$\epsilon^2$ & $\epsilon_1^2+\epsilon_2^2+\epsilon_3^2$ \\
$\alpha^2$ & $4b^2\epsilon_1^2+(\epsilon^2 - b^2)^2$ \\
$\alpha_\pm^2$ & $\alpha \pm (\epsilon^2 - b^2)$ \\
$\chi_+$ & $\cosh( \alpha_+ q\Delta s/\sqrt{2}mc)$ \\
$\chi_-$ & $\cos( \alpha_- q\Delta s/\sqrt{2}mc)$ \\
$\sigma_+$ & $\sinh( \alpha_+ q\Delta s /\sqrt{2}mc)/\alpha_+$ \\
$\sigma_-$ & $\sin( \alpha_- q\Delta s /\sqrt{2}mc)/\alpha_-$ \\
$\Xi_\pm$ & $(\chi_+ \pm \chi_-)/2$ \\
$\Sigma_\pm$ & $(\sigma_+ \pm \sigma_-)/2$ \\
$F$ & $b^2+\epsilon^2$ \\
$G$ & $-b^2-\epsilon_1^2+\epsilon_2^2+\epsilon_3^2$ \\
$H$ & $b^2+\epsilon_1^2-\epsilon_2^2+\epsilon_3^2$ \\
$J$ & $b^2+\epsilon_1^2+\epsilon_2^2-\epsilon_3^2$ \\
\hline
\end{tabular}
\end{center}
\end{table}

\begin{figure}
\includegraphics[width=3.0in]{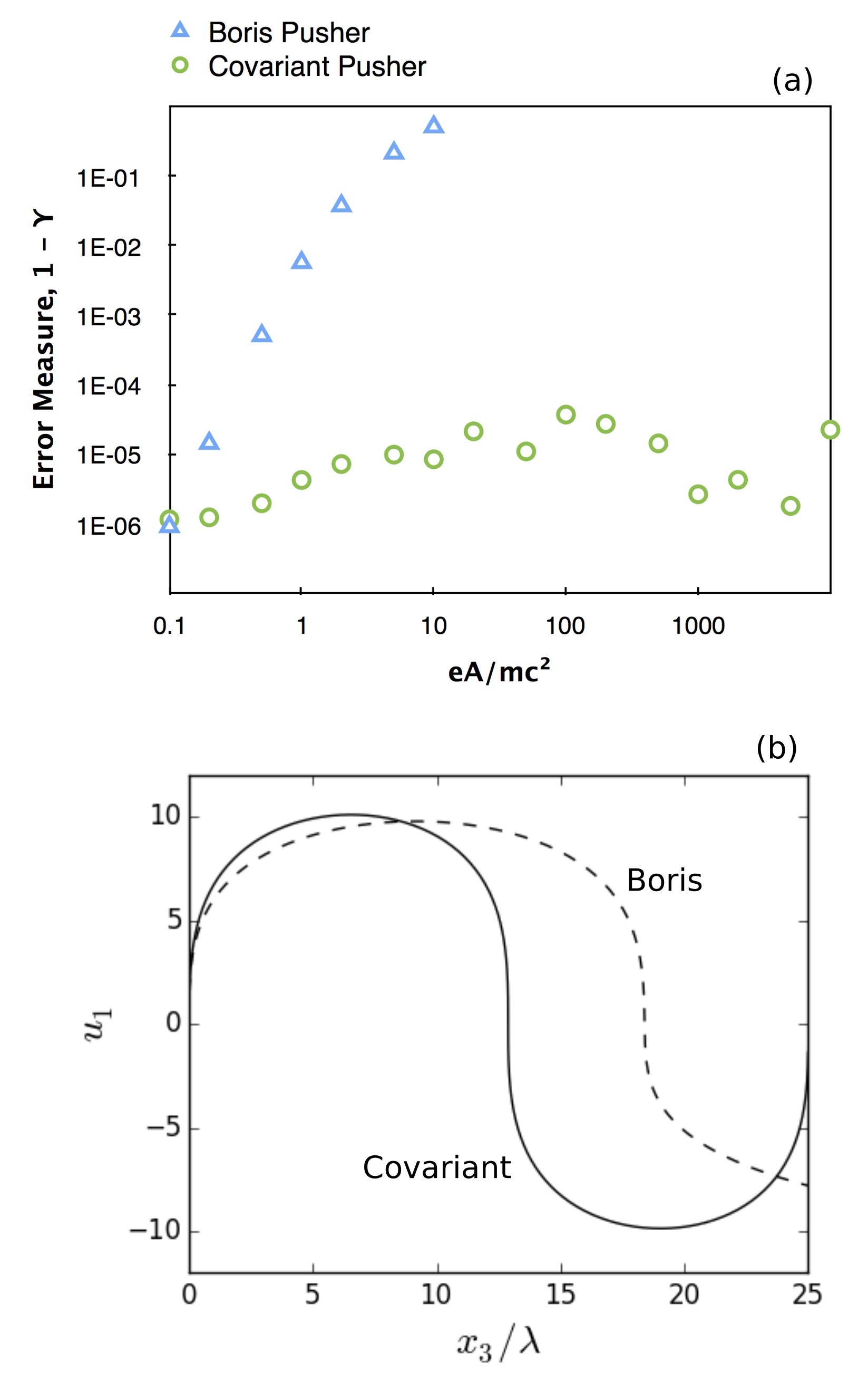}
\caption{Errors due to the Boris pusher and the covariant pusher introduced in this paper.  The error measure, $1-\Upsilon$, is plotted as a function of $a$ in (a), and the orbits corresponding to the case $a=10$ are plotted in (b).}
\label{fig:accuracy}
\end{figure}

In order to emphasize the importance of the covariant pusher just described, its accuracy is compared with that of the Boris pusher \cite{boris70} in Fig.~\ref{fig:accuracy}.  The simulation initializes a single electron into a monochromatic plane wave such that $\Upsilon = 1$ and $u_1 = -qA_1/mc^2$.  The error measure $1-\Upsilon$, which should be zero at all times, is plotted in Fig.~\ref{fig:accuracy}(a). The accuracy of the Boris pusher suffers with increasing $a$, while the covariant pusher is accurate for any $a$.  Phase space orbits for $a=10$ are plotted in Fig.~\ref{fig:accuracy}(b).  The covariant pusher correctly recovers the spatial period, which is known analytically to be $a^2\lambda/4$.  In contrast, the Boris pusher gives a spuriously long spatial period.

It is important to note that the covariant pusher gives accurate results much faster than the Boris pusher.  The time step used in Fig.~\ref{fig:accuracy} is $2\pi/267\omega$.  In the case of the Boris pusher, the time step is measured in the lab frame, whereas in the case of the covariant pusher, it is measured in the frame of the electron.  The time elapsed in the lab frame is always longer than the time elapsed according to the electron's own clock.  As a result, the covariant pusher always requires fewer steps than the Boris pusher.  This is related to the fact that the phase step in the covariant pusher is constant, while the phase step in the Boris pusher decreases with increasing forward momentum.  As an example, in Fig.~\ref{fig:accuracy}(b), the covariant pusher curve is generated in 267 steps, while the Boris pusher curve is generated in 6800 steps.

It is natural to ask whether using an adaptive time step is an alternative to using the unsplit matrix exponential $\Lambda$.  This approach is elaborated upon in \cite{gordon15.NRLMR}, where the pusher is in covariant form, but retains the conventional time step splitting.  This turns out to be far less efficient than using the unsplit $\Lambda$.  

Another advantage of expressing the particle pusher in covariant fashion is that the Lorentz-Abraham-Dirac formula for the radiation reaction force takes the simple form
\be
R = \frac{2q^2}{3mc}\left(\frac{d^2u}{ds^2} - u u^T g \frac{d^2u}{ds^2}\right)
\ee
The Landau and Lifshitz (LL) formula \cite{landau-ctf} is derived by substituting for $d^2u/ds^2$ the value obtained in the absence of radiation reaction.  In three dimensional notation, the LL formula is extremely unwieldy, and even in four dimensional form, it {\em appears} to require expensive evaulations of all spacetime derivatives of the field tensor.  When the covariant pusher described above is used, a simple and elegant alternative becomes readily available.  Namely, by splitting each step into two half-steps, $d^2u/ds^2$ can be evaluated by direct finite differencing.  That is, during each step generate
\be
u^{(1)} = \Lambda\left(s+\frac{\Delta s}{4},\frac{\Delta s}{2}\right) u^{(0)}
\ee 
\be
u^{(2)} = \Lambda\left(s+\frac{3\Delta s}{4},\frac{\Delta s}{2}\right) u^{(1)}
\ee 
This requires field evaluations at only two spacetime points.  Now, in the absence of radiation reaction, $u^{(0)} = u(s)$, $u^{(1)} = u(s+\Delta s/2)$, and $u^{(2)} = u(s+\Delta s)$.  The LL reaction force is therefore given by the matrix equation
\be
R = \frac{2q^2}{3mc}\left(\delta^2u - u u^T g \delta^2u\right)
\ee
where
\be
\delta^2u = \frac{u^{(2)}-2u^{(1)}+u^{(0)}}{\Delta s^2/4}
\ee
is the finite difference form of $d^2u/ds^2$.  The updated four-velocity, including the LL reaction force, is
\be
u(s+\Delta s) = u^{(2)} + R\Delta s
\ee
Here, a simple Euler advance is justified by the expectation that the reaction force is small.

\section*{Appendix B: Elaboration on the Two-Step Model}

The quantum mechanical $S$-matrix is at the foundation of the SFA.  It has a simple interpretation as the matrix of amplitudes of all possible transitions between in and out states.  The $S$-matrix viewpoint holds that the electron dynamics during the interaction are not observable.  The $S$-matrix expresses this fundamental ignorance by giving only the transition amplitude between states in the distant past or future, when the interaction is negligible.  In contrast, the two-step model allows for a full dynamical description of the electron motion after a certain critical point in time, which is loosely identified with the ``time of ionization.''  This raises a problem of interpretation due to the fact that time is not a quantum mechanical observable.

Before addressing the resolution of this problem, it is useful to note that many quantum mechanical $S$-matrix calculations do in fact introduce something like a critical time in the course of making mathematical approximations.  In particular, the total $S$-matrix amplitude is obtained from the saddle point method, which introduces a summation over saddle points, each of which corresponds to a particular phase of the applied field.  There are two saddle points per optical cycle.  One is tempted to interpret this procedure as a summation over ionization events, which are localized in phase near the peaks of the field.

In order to interpret the time of ionization, Bohmian mechanics is used.  Bohmian mechanics is equivalent to quantum mechanics in all its statistical predictions, but treats the electron coordinate as a hidden variable \cite{bohm-hiley}.  The electron responds to electromagnetic forces classically, but also responds to a quantum force that depends on the wavefunction.  In this context, the time of ionization is easy to define.  Let $t_c$ be the earliest time such that for all $t>t_c$, the classical force dominates the quantum force.  Then $t_c$ is the critical time after which the electron behaves classically.  This is as useful a notion of ``time of ionization'' as one is likely to discover.

In the Bohmian picture, the coordinates of an electron in a stationary state can take a continuous range of definite values.  The statistical distribution of these initial coordinates determines all the subsequent statistics, such as the distribution of ionization times.  In order to proceed, it is not necessary to solve the whole Bohmian mechanics problem.  Instead, a single hidden variable, $H$, is introduced into the initial electron distribution.  It turns out that if the electron distribution contains a factor $e^{-H}$, and if $t_c$ is defined by
\be
\int_{-\infty}^{t_c} w(t') dt' = H
\ee
then $w(t)$ is the ionization rate \cite{gordon15.NRLMR}.  Therefore, $H$ is a hidden variable that recovers any quantum mechanical ionization rate, $w(t)$, that one chooses to use.  Moreover, the time $t_c$ is the time of ionization, after which classical particle tracking is appropriate.  What is missing is a clear specification of the electron coordinates at the moment of ionization, $x(t_c)$ and $u(t_c)$.  In order to rigorously specify these, knowledge of all possible Bohmian trajectories would be needed.  An interesting task for future research would be to utilize Bohmian trajectories, computed numerically \cite{gordon15}, as inputs into the two-step model.  In this work, appeal is made to the ITM, which suggests that $u(t_c) \approx (1,0,0,0)^T$.  The spatial coordinate of the ionized electron is assumed to coincide with that of the parent ion.

In carrying out the classical particle tracking, expressions for the laser field components are needed as inputs into the particle pusher algorithm.  Let $\psi$ be the lowest order Hermite-Gaussian solution of the scalar paraxial wave equation \cite{Siegman}.  In order to impose a temporal envelope, let $A_1 = \psi\exp[-(z-vt)^2/c^2\tau^2]$, where $v$ is the velocity of the envelope, taken to be $c$ in this work.  Then, to lowest order in $\lambda/r_0$, $A_3$ is found from $\nabla \cdot {\bf A} = 0$.  As usual, $A_2$ is neglected.  The field tensor is obtained by differentiating $A_\mu$, also to lowest order in $\lambda/r_0$.

Finally, we briefly comment on the electron spin.  The spin of the electron is accounted for in the SFA rate law, which gives the spin averaged ionization rate \cite{klaiber13}.  Spin effects in the context of the SFA are explicitly considered in Ref.~\cite{yakaboylu15}.  In the two-step model, spin effects are neglected for $t>t_c$.  It is possible to isolate the spin terms in the Dirac equation \cite{Landau-qed}, and form the ratio of spin terms to other terms.  In a laser field, this ratio is of order $\hbar\omega/amc^2$, which is miniscule for the parameters considered in this paper.

\newpage

\bibliographystyle{unsrt}

\begin{thebibliography}{10}

\bibitem{javanainen88}
J.~Javanainen, J.H. Eberly, and Q.~Su.
\newblock Numerical simulations of multiphoton ionization and above-threshold
  electron spectra.
\newblock {\em Phys. Rev. A}, 38(7):3430--3446, 1988.

\bibitem{corkum89}
P.~B. Corkum, N.~H. Burnett, and F.~Brunel.
\newblock Above-threshold ionization in the long-wavelength limit.
\newblock {\em Phys. Rev. Lett.}, 62(11):1259--1262, 1989.

\bibitem{moore99}
C.I. Moore, A.~Ting, S.~Mc{N}aught, J.~Qiu, H.R. Burris, and P.~Sprangle.
\newblock A laser-accelerator injector based on laser ionization and
  ponderomotive acceleration of electrons.
\newblock {\em Phys. Rev. Lett.}, 82(8):1688--1691, 1999.

\bibitem{moore01.lipa}
C.~I. Moore, A.~Ting, T.~Jones, E.~Briscoe, B.~Hafizi, R.~F. Hubbard, and
  P.~Sprangle.
\newblock Measurements of energetic electrons from the high-intensity laser
  ionization of gases.
\newblock {\em Phys. Plasmas}, 8(5):2481, May 2001.

\bibitem{keldysh65}
L.~V. Keldysh.
\newblock Ionization in the field of a strong electromagnetic wave.
\newblock {\em Soviet Physics JETP}, 20(5):1307--14, May 1965.

\bibitem{klaiber13}
M.~Klaiber, E.~Yakaboylu, and K.Z. Hatsagortsyan.
\newblock Above-threshold ionization with highly charged ions in superstrong
  laser fields. {II}. {R}elativistic {C}oulomb-corrected strong-field
  approximation.
\newblock {\em Phys. Rev. A}, 87:023418--1--023418--11, 2013.

\bibitem{yudin01}
G.L. Yudin and M.Y. Ivanov.
\newblock Nonadiabatic tunnel ionization: {L}ooking inside a laser cycle.
\newblock {\em Phys. Rev. A}, 64:013409--1--013409--4, Jun 2001.

\bibitem{popov2004}
V.S. Popov.
\newblock Tunnel and multiphoton ionization of atoms and ions in a strong laser
  field ({K}eldysh theory).
\newblock {\em Physics-Uspekhi}, 47(9):855--885, 2004.

\bibitem{ELI-NP}
O.~Tesileanu, D.~Ursescu, R.~Dabu, and N.V. Zamfir.
\newblock Extreme light infrastructure - nuclear physics.
\newblock {\em J. Phys.: Conf. Series}, 420:012157--1--012157--7, 2013.

\bibitem{kimura95}
W.D. Kimura, G.H. Kim, R.D. Romea, L.C. Steinhauer, I.V. Pogorelsky, K.P.
  Kusche, R.C. Fernow, X.~Wang, and Y.~Liu.
\newblock Laser acceleration of relativistic electrons using the inverse
  {C}herenkov effect.
\newblock {\em Phys. Rev. Lett.}, 74(4):546--549, Jan 1995.

\bibitem{hartemann98}
F.V. Hartemann, J.R.~Van Meter, A.L. Troha, E.C. Landahl, N.C.~Luhmann Jr.,
  H.A. Baldis, A.~Gupta, and A.K. Kerman.
\newblock Three-dimensional relativistic electron scattering in an
  ultrahigh-intensity laser focus.
\newblock {\em Phys. Rev. E}, 58(4):5001--5012, Oct 1998.

\bibitem{wang01}
P.X. Wang, Y.K. Ho, X.Q. Yuan, Q.~Kong, N.~Cao, A.M. Sessler, E.~Esarey, and
  Y.~Nishida.
\newblock Vacuum electron acceleration by an intense laser.
\newblock {\em Appl. Phys. Lett.}, 78(15):2253--2255, Apr 2001.

\bibitem{gupta07}
D.N. Gupta, N.~Kant, D.E. Kim, and H.~Suk.
\newblock Electron acceleration to {G}e{V} energy by a radially polarized
  laser.
\newblock {\em Phys. Lett. A}, 368:402--407, 2007.

\bibitem{cline13}
D.~Cline, L.~Shao, X.~Ding, Y.~Ho, Q.~Kong, and P.~Wang.
\newblock First observation of acceleration of electrons by a laser in a
  vacuum.
\newblock {\em J. Modern Phys.}, 4:1--6, 2013.

\bibitem{palmer87}
R.B. Palmer.
\newblock An introduction to acceleration mechanisms.
\newblock Technical Report {SLAC}-{PUB}-4320, Stanford Linear Accelerator
  Center, May 1987.

\bibitem{reiss92}
H.R. Reiss.
\newblock Theoretical methods in quantum optics: S-matrix and {K}eldysh
  techniques for strong-field processes.
\newblock {\em Prog. Quant. Electr.}, 16:1, 1992.

\bibitem{hubbard02}
R.F. Hubbard, B.~Hafizi, A.~Ting, D.~Kaganovich, P.~Sprangle, and A.~Zigler.
\newblock High intensity focusing of laser pulses using a short plasma channel
  lens.
\newblock {\em Phys. Plasmas}, 9(4):1431--1442, Apr 2002.

\bibitem{katzir09}
Y.~Katzir, S.~Eisenmann, Y.~Ferber, A.~Zigler, and R.F. Hubbard.
\newblock A plasma microlens for ultrashort high power lasers.
\newblock {\em Appl. Phys. Lett.}, 95(3):031101--1--031101--3, 2009.

\bibitem{rohrlich02}
F.~Rohrlich.
\newblock Dynamics of a classical quasi-point charge.
\newblock {\em Phys. Lett. A}, 303:307--310, 2002.

\bibitem{hartemann05}
F.V. Hartemann, D.J. Gibson, and A.K. Kerman.
\newblock Classical theory of {C}ompton scattering: {A}ssessing the validity of
  the {D}irac-{L}orentz equation.
\newblock {\em Phys. Rev. E}, 72:026502--1--026502--9, 2005.

\bibitem{dipiazza09}
A.~Di Piazza, K.Z. Hatsagortsyan, and C.H. Keitel.
\newblock Strong signatures of radiation reaction below the radiation-dominated
  regime.
\newblock {\em Phys. Rev. Lett.}, 102:254802--1--254802--4, 2009.

\bibitem{thomas12}
A.G.R. Thomas, C.P. Ridgers, S.S. Bulanov, B.J. Griffin, and S.P.D. Mangles.
\newblock Strong radiation-damping effects in a gamma-ray source generated by
  the interaction of a high-intensity laser with a wakefield-accelerated
  electron beam.
\newblock {\em Phys. Rev. X}, 2:041004--1--041004--13, 2012.

\bibitem{kumar13}
N.~Kumar, K.Z. Hatsagortsyan, and C.H. Keitel.
\newblock Radiation-reaction-force-induced nonlinear mixing of {R}aman
  sidebands of an ultraintense laser pulse in a plasma.
\newblock {\em Phys. Rev. Lett.}, 111:105001--1--105001--5, 2013.

\bibitem{kravets13}
Y.~Kravets, A.~Noble, and D.~Jaroszynski.
\newblock Radiation reaction effects on the interaction of an electron with an
  intense laser pulse.
\newblock {\em Phys. Rev. E}, 88:011201--1--011201--5, 2013.

\bibitem{gordon15.NRLMR}
D.F. Gordon, J.P. Palastro, B.~Hafizi, D.~Kaganovich, L.~Johnson, R.~Hubbard,
  M.H. Helle, and A.~Ting.
\newblock x{LIPA}: promotion of electrons from the {K}-shell to 2 {G}e{V} using
  10 {PW} laser pulses.
\newblock Technical Report NRL/MR/6791--15-9634, Naval Research Laboratory,
  Washington, DC, Aug 2015.

\bibitem{boris70}
J.P. Boris.
\newblock Relativistic plasma simulation-optimization of a hybrid code.
\newblock In Proc. 4th Conf. Numer. Simul. Plasmas, 1970.

\bibitem{verboncoeur05}
J.~P. Verboncoeur.
\newblock particle simulation of plasmas: review and advances.
\newblock {\em Plasma Phys. and Control. Fusion}, 47:A231--A260, 2005.

\bibitem{landau-ctf}
L.D. Landau and E.M. Lifschitz.
\newblock {\em Classical Theory of Fields}.
\newblock Pergamon Press, Oxford, England, 1980.

\bibitem{bohm-hiley}
D.~Bohm and B.J. Hiley.
\newblock {\em The undivided universe}.
\newblock Routledge, 270 Madison Ave., New York NY 10016, 1993.

\bibitem{gordon15}
D.F. Gordon, B.~Hafizi, and A.S. Landsman.
\newblock Amplitude flux, probability flux, and gauge invariance in the finite
  volume scheme for the schroedinger equation.
\newblock {\em J. Comp. Phys.}, 280:457--464, Jan 2015.

\bibitem{Siegman}
Anthony~E. Siegman.
\newblock {\em Lasers}.
\newblock University Science Books, Mill Valley, California, 1986.

\bibitem{yakaboylu15}
E.~Yakaboylu, M.~Klaiber, and K.Z. Hatsagortsyan.
\newblock Above threshold ionization with highly charged ions in superstrong
  laser fields. iii. spin effects and their dependence on laser polarization.
\newblock {\em Phys. Rev. A}, 91:063407, Jun 2015.

\bibitem{Landau-qed}
V.B. Berestetsky, E.M. Lifshitz, and L.P. Pitaevskii.
\newblock {\em Quantum Electrodynamics}.
\newblock Pergamon Press, 1980.

\end{thebibliography}

 \end{document}